\documentclass[12pt]{article}
\usepackage{amssymb,amsmath,epsfig}

\begin{document}

\title{\bf Holographic dark energy reconstruction in $f(T,\mathcal{T})$ gravity}
\author{Ines. G. Salako$^a$ \thanks{inessalako@gmail.com},
Abdul Jawad$^b$ \thanks{jawadab181@yahoo.com;
abduljawad@ciitlahore.edu.pk} and
Surajit Chattopadhyay$^c$ \thanks{surajcha@iucaa.ernet.in}\\
$^{a}$Institut de Math\'{e}matiques et de Sciences Physiques (IMSP)
\\ 01 BP 613 Porto-Novo, B\'{e}nin.\\
$^b$Department of Mathematics, COMSATS Institute of\\ Information
Technology, Lahore-54000, Pakistan.\\
$^c$Pailan College of Management and Technology, Bengal\\
Pailan Park, Kolkata-700 104, India.}

\date{}
\maketitle

\begin{abstract}
The present paper reports a holographic reconstruction scheme for
$f(T,\mathcal T)$ gravity proposed in Harko et al. $\emph{JCAP}\;
12(2014)021$ where $T$ is the torsion scalar and $\mathcal{T}$ is
the trace of the energy-momentum tensor considering future event
horizon as the enveloping horizon of the universe. We have
considered $f(T, \mathcal T)=T + \gamma g(\mathcal T)$ and
$f(T,\mathcal T) =\beta \mathcal T + g(T)$ for reconstruction. We
observe that the derived $f(T,\mathcal T)$ models can represent
phantom or quintessence regimes of the universe which are compatible
with the current observational data.
\end{abstract}

\maketitle


\section{Introduction}\label{sec1}

Accelerated expansion of the current universe, which is suggested by
the two independent observational signals on distant Type Ia
Supernovae (SNeIa) (Riess et al. 1998; Perlmutter 1999; Knop et al.
2003), the Cosmic Microwave Background (CMB) temperature
anisotropies measured by the WMAP and Planck satellites (Spergel et
al. 2003; Komatsu et al. 2011; Ade et al. 2013) and Baryon Acoustic
Oscillations (BAO) (Eisenstein et al. 2005; Percival et al. 2010),
is well documented in the literature (Shafieloo et al. 2009;
Copeland et al. 2006) and the search for causes behind this
accelerated expansion falls into two representative categories: in
the first, the concept of ``dark energy" (DE) is introduced  in the
right-hand side of the Einstein equation in the framework of general
relativity (for good reviews see (Copeland et al. 2006; Bamba et al.
2012; Caldwell and Kamionkowski 2009)) while in the second one the
left-hand side of the Einstein equation is modified, leading to a
modified gravitational theory (which is well reviewed in (Nojiri and
Odintsov 2011; Clifton et al. 2012; Capozziello et al. 2012;
Tsujikawa 2010)).

It was demonstrated in a recent review by Bamba et al. (2012) that
both DE models and modified gravity theories are in agreement with
data and hence, these two rival approaches could not be
discriminated, unless higher precision probes of the expansion rate
and the growth of structure of the universe will be available.
Origin of DE is one of the most serious problems in modern cosmology
(Tsujikawa 2010; Sahni and Starobinsky 2000; Carroll 2001;
Padmanabhan 2003; Peebles and Ratra 2003; Sahni et al. 2008). The
simplest candidate for DE is cosmological constant $\Lambda$, which
is extensively reviewed in (Peebles and Ratra 2003). The dynamical
DE models can be distinguished from the cosmological constant by
considering the evolution of the equation of state (EoS) parameter
$w_{DE}=p_{DE}/\rho_{DE}$, where $p_{DE}$ is the pressure and
$\rho_{DE}$ is the density of DE. Various candidates of DE are
proposed till date that do not involve the cosmological constant.
Although the current observational data are not sufficient to
provide some preference of other DE models over the $\ Lambda$CDM
model, it says a nothing about the time evolution of the EoS
parameter. Dynamic DE models proposed so far include scalar-field
models of DE (Amendola 2000; de la Macorra and Filobello 2008; Forte
2004; Singh et al. 2003; Kunz and Sapone 2006; Novosyadlyj et al.
2013), k-essence (Malquarti et al. 2003; Bilic 2008) and Chaplygin
gas (Gorini et al. 2008; Pun et al. 2008; Setare 2007). There is
another model of DE that is based on the holographic principle
according to which the entropy of a system scales not with its
volume but with its surface area. This DE candidate is dubbed as
holographic DE (Elizalde et al. 2005; Nojiri and Odintsov 2006a; del
Campo et al. 2011; Cui and Zhang 2014; Huang and Gong 2004; Huang
and Li 2005; Zhang and Wu 2007).

Now we come to the other approach towards the accelerated expansion
of the universe i.e. the modified gravity theory". The current
contribution being aimed at exploring a cosmological reconstruction
in the framework of a modified gravity theory, let us have a brief
overview of the theories of modified gravity as the current
contribution is going. Nowadays, modified gravity has become a
crucial part of theoretical cosmology (Nojiri and Odintsov 2007a;
Nojiri and Odintsov 2007b; Bamba et al. 2012). It is proposed as
generalization of General Relativity with the purpose to understand
the qualitative change of gravitational interaction in the very
early and/or very late universe. In particular, modified gravity not
only describes the early-time inflation and late-time acceleration
but also proposes the unified consistent description of the universe
evolution epochs sequence: inflation, radiation/matter dominance and
DE (Nojiri and Odintsov 2014). Nojiri and Odintsov (2007b)
summarized the usefulness of modified gravity as follows:
\begin{enumerate}
\item it provides natural gravitational alternative
for DE.
\item it presents very natural unification of the early-time inflation and late-time acceleration thanks
to different role of gravitational terms relevant at small and at
large curvature.
\item it may serve as the basis for unified explanation of DE and dark matter.
\end{enumerate}
Reviews on modified gravity include (Bamba et al. 2012; Clifton et
al. 2012; Nojiri and Odintsov 2007b; De Felice and Tsujikawa 2010).
One of the simplest modifications to the general relativity is the
$f(R)$ gravity in which the Lagrangian density $f$ is an arbitrary
function of Ricci scalar $R$ (De Felice and Tsujikawa 2010). The
$f(R)$ gravity has been reviewed in De Felice and Tsujikawa (2010).
The model with $f(R)=R+\alpha R^2$ with $\alpha>0$, proposed by
Starobinsky (1980), can lead to accelerated expansion of the
universe. DE models based on $f(R)$ theories have been extensively
studied as the simplest modified gravity scenario to realize the
late-time acceleration (Capozziello 2002; Capozzielloa et al. 2006;
Nojiri and Odintsov 2011; Cognola et al. 2005; Nojiri and Odintsov
2007c; Nojiri and Odintsov 2008). A generalization of $f(R)$
modified theories of gravity was proposed in (Bertolami et al. 2007)
by coupling an arbitrary function of the Ricci scalar $R$ with the
matter Lagrangian density $L_m$ (Poplawski 2006). Nevertheless,
other kinds of theories have been suggested which include other
curvature invariants, such as the Gauss-Bonnet gravity (Bamba et al.
2014), $f(G)$ gravity (Zhao et al.  2012; Daouda et al. 2012; Salako
et al. 2013; Rodrigues et al. 2014), $f(T)$ gravity ($T$ is torsion)
(Myrzakulov 2011), Horava-Lifshitz cosmology (Kiritsis and Kofinas
2009), Brans-Dicke cosmology (Lee et al. 2011) etc.

In a recent work, Harko et al. (2014) presented an extension of
$f(T)$ gravity, allowing for a general coupling of the torsion
scalar $T$ with the trace of the matter energy-momentum tensor
$\mathcal T$ that resulted in a new modified gravity dubbed $f(T,
\mathcal T)$ gravity that helps one obtaining unified description of
the initial inflationary phase, the subsequent non-accelerating,
matter-dominated expansion, and then the transition to a late-time
accelerating phase. Subsequently (Momeni and Myrzakulov 2014; Junior
et al. 2015) showed how $f(T, \mathcal T)$ can be reconstructed for
$\Lambda$CDM as the most popular and consistent model. The present
paper reports a holographic reconstruction scheme for $f(T, \mathcal
T)$ using the models $f(T, \mathcal T)=T+\gamma g(\mathcal T)$ and
$f(T,\mathcal T)=\beta \mathcal T+g(T)$. Rest of the paper has the
following sequence: We present a general introduction of the $f(T,
\mathcal T)$ gravity in the next section. We discuss the features of
HDE and the corresponding reconstruction scheme for the aforesaid
models in section \textbf{4}. We conclude our results in the last
section.

\section{ $f(T,\mathcal{T})$ Gravity}

The line element is defined as follows
\begin{eqnarray}
ds^2=g_{\mu\nu}dx^\mu dx^\nu=\eta_{ij}\theta^i\theta^j\,,
\end{eqnarray}
with the definition
\begin{eqnarray}
d^\mu=e_{i}^{\;\;\mu}\theta^{i}; \,\quad
\theta^{i}=e^{i}_{\;\;\mu}dx^{\mu}.
\end{eqnarray}
Here, $\eta_{ij}=diag(1,-1,-1,-1)$ (Minkowskian metric) while the
tetrad components $\{e^{i}_{\;\mu}\}$ satisfy the following
relations
\begin{eqnarray}
e^{\;\;\mu}_{i}e^{i}_{\;\;\nu}=\delta^{\mu}_{\nu},\quad
e^{\;\;i}_{\mu}e^{\mu}_{\;\;j}=\delta^{i}_{j}.
\end{eqnarray}
The Levi-Civita connection (in general relativity) has the form
\begin{equation}
\overset{\circ }{\Gamma }{}_{\;\;\mu \nu }^{\rho } =
\frac{1}{2}g^{\rho \sigma }\left(
\partial _{\nu} g_{\sigma \mu}+\partial _{\mu}g_{\sigma \nu}-\partial _{\sigma}g_{\mu \nu}\right)\;,
\end{equation}%
which exists for nonzero spacetime curvature but zero torsion. On
the other hand, teleparallel theory and its modified versions
contains Weitzenbock's connection which has the following form
\begin{eqnarray}
\Gamma^{\lambda}_{\mu\nu}=e^{\;\;\lambda}_{i}\partial_{\mu}
e^{i}_{\;\;\nu}=-e^{i}_{\;\;\mu}\partial_\nu e_{i}^{\;\;\lambda}.
\end{eqnarray}
This connection has main geometrical objects and torsion is one of
them, which is defined as
\begin{eqnarray}
T^{\lambda}_{\;\;\;\mu\nu}=
\Gamma^{\lambda}_{\mu\nu}-\Gamma^{\lambda}_{\nu\mu},
\end{eqnarray}
and the corresponding contorsion tensor is
\begin{equation}\label{K}
K_{\;\;\mu \nu }^{\lambda} \equiv \widetilde{\Gamma} _{\;\mu \nu
}^{\lambda } -\overset{\circ}{\Gamma }{}_{\;\mu \nu
}^{\lambda}=\frac{1}{2}(T_{\mu }{}^{\lambda}{}_{\nu } +
T_{\nu}{}^{\lambda }{}_{\mu }-T_{\;\;\mu \nu }^{\lambda})\;,
\end{equation}
and its other from is
\begin{eqnarray}
K^{\mu\nu}_{\;\;\;\;\lambda}=-\frac{1}{2}
\left(T^{\mu\nu}_{\;\;\;\lambda}-T^{\nu\mu}_{\;\;\;\;\lambda}+T^{\;\;\;\nu\mu}_{\lambda}\right)\,\,.
\end{eqnarray}
The torsion and contorsion help us in defining the new tensor
$S_{\lambda}^{\;\;\mu\nu}$ as follows
\begin{eqnarray}
S_{\lambda}^{\;\;\mu\nu}=\frac{1}{2}\left(K^{\mu\nu}_{\;\;\;\;\lambda}+
\delta^{\mu}_{\lambda}T^{\alpha\nu}_{\;\;\;\;\alpha}-
\delta^{\nu}_{\lambda}T^{\alpha\mu}_{\;\;\;\;\alpha}\right).\label{S}
\end{eqnarray}
Also, one can defined the torsion scalar as follows
\begin{eqnarray}
T=T^{\lambda}_{\;\;\;\mu\nu}S^{\;\;\;\mu\nu}_{\lambda}
\end{eqnarray}

Since we are dealing with a modified version of the teleparallel
gravity whose action can be written as
\begin{eqnarray}
 S= \int e \left[\frac{T+f(T,\mathcal{T})}{2\kappa^2} +\mathcal{L}_{m} \right]d^{4}x   \label{eq9}
\end{eqnarray}
where $\kappa^{2} = 8 \pi G $. The variation of this action
(\ref{eq9}) according to tetrads gives (Momeni and Myrzakulov 2014;
Junior et al. 2015; Harko et al. 2014)
\begin{eqnarray}
&& S^{\;\;\; \nu \rho}_{\mu}\;f_{TT}\; \partial_{\rho} T   +
\Big[e^{-1} e^{i}_{\;\; \mu}\partial_{\rho} \Big(e e^{\;\;
\mu}_{i}S^{\;\;\; \nu\lambda}_{\alpha} \Big) +T^{\alpha}_{\;\;\;
\lambda \mu}   S^{\;\;\; \nu \lambda}_{\alpha} \Big] \Big(1 +
f_{T}\Big)+\cr && \frac{1}{4}\delta^{\nu}_{\mu} T  =S^{\;\;\; \nu
\rho}_{\mu}\; f_{T\mathcal{T}}\; \partial_{\rho} \mathcal{T} +
f_{\mathcal{T}}\;\Big(\frac{\Theta^{\nu}_{\mu}
+\delta^{\nu}_{\mu}\;p }{2}\Big) -\frac{1}{4}\delta^{\nu}_{\mu}
f(\mathcal{T}) +\frac{\kappa^{2}}{2} \Theta^{\nu}_{\mu}
\label{lagran1}
\end{eqnarray}
where  $f_{T} = \partial f/\partial T$, $f_{T\mathcal{T}}  =
\partial^{2}f/\partial T\partial \mathcal{T}$, $f_{TT}  =
\partial^{2}f/\partial T^{2}$ and $\Theta^{\nu}_{\mu}$ is the
energy-momentum tensor of the matter fields. Here we study the
cosmological dynamics of the present modified gravity in flat FRW
universe. The FRW metric is defined as
\begin{eqnarray}
ds^{2}= dt^{2} - a^{2}(t)\left(dx^2+dy^2+dz^2\right).  \label{eq10}
\end{eqnarray}
We use diagonal tetrads $\{e^{a}_{\;\; \mu}\}= diag[1,a,a,a]$ and
its determinant is $a^{3}$. The torsion and contorsion tensors has
following non-zero components
\begin{eqnarray}
T^{1}_{\;\;\; 01}= T^{2}_{\;\;\; 02}=T^{3}_{\;\;\;
03}=\frac{\dot{a}}{a},\quad K^{01}_{\;\;\;\;1}=K^{
02}_{\;\;\;\;2}=K^{ 03}_{\;\;\;\;3}= \frac{\dot{a}}{a}, \label{eq12}
\end{eqnarray}
and the components of the tensor $S^{\;\;\; \mu\nu}_{\alpha}$  are
\begin{eqnarray}
S^{\;\;\; 11}_{0}=S^{\;\;\; 22}_{0}=S^{\;\;\;
33}_{0}=\frac{\dot{a}}{a}.
\end{eqnarray}
However, the torsion scalar has the following form
\begin{eqnarray}
 T= -6H^{2}, \label{manuel8}
\end{eqnarray}
in the present scenario and $H=\dot{a}/a$ denotes the Hubble
parameter. We mention that the expression of the trace of matter
energy-momentum tensor $\Theta =\mathcal{T}= (\rho_m-3 p_m)$. We
consider the ordinary DM whose EoS is $p_{m} = \omega_{m} \rho_{m}$
and the corresponding energy-momentum tensor is
\begin{eqnarray}
\Theta ^{\nu}_{\mu} = diag(1,-\omega_{m}, -\omega_{m}, -\omega_{m} )
\rho_m.
\end{eqnarray}

\section{Reconstruction of Holographic $f(T,\mathcal{T})$ Dark Energy}

\subsection{Holographic Dark Energy}

In this section, we present a general formalism of HDE density. The
density $\rho_{DE}$ can be written as (Wu and Zhu 2008;  Houndjo and
Piattella 2011; Setare and Darabi 2011)
\begin{eqnarray}\label{manuel12}
\rho_{DE}=\frac{3e^2}{R^2_h},\quad
R_h=a(t)\int^{\infty}_{t}\frac{d\tilde{t}}{a(\tilde{t})}=a\int^{\infty}_{a}\frac{da}{Ha^2}
\end{eqnarray}
where $e$ is a constant. By using of the critical energy density
$\rho_{cr}=3H^2$, we can define the dimensionless DE as
\begin{eqnarray}\label{manuel15}
\Omega_{DE}=\frac{\rho_{DE}}{\rho_{cr}}=\frac{e^2}{H^2R^2_h}\,\,\,.
\end{eqnarray}
Using the definitions of $\Omega_{DE}$ and $\rho_{cr}$, we get
\begin{eqnarray}
\dot{R}_h&=&HR_h-1=\frac{e}{\sqrt{\Omega_{DE}}}-1.\label{manuel16}
\end{eqnarray}
The continuity equation for DE becomes
\begin{eqnarray}\label{manuel17}
\dot{\rho}_{DE}+3H\left(\rho_{DE} + p_{DE}\right)=0\,\,\,.
\end{eqnarray}
In terms of fractional energy density, the time rate of the HDE
density becomes
\begin{eqnarray}\label{manuel18}
\dot{\rho}_{DE}=-\frac{-2}{R_h}\left(\frac{e}{\sqrt{\Omega_{DE}}}-1\right)\rho_{DE}\,\,\,,
\end{eqnarray}
from which, using (\ref{manuel17}), we get
\begin{eqnarray}\label{manuel19}
\omega_{DE}=-\left(\frac{1}{3}+\frac{2\sqrt{\Omega_{DE}}}{3e}\right)\,\,\,.
\end{eqnarray}
From the above equation, one can analyze the behavior of EoS
parameter for $\Omega_{DE} \rightarrow 1$ (in the future) as
follows: $\omega_{DE}>-1$ and behaves like a quintessence for $e>1$.
While the universe approaches to de Sitter phase for $e=1$ and it
enters into phantom phase (with $\omega_{DE}<-1$) for $e<1$. Thus,
the parameter $e$ plays a crucial role in the evolution of the
universe through HDE.

By assuming two particular actions of $f(T,\mathcal{T})$ Lagrangian,
we reconstruct HDE $f(T,\mathcal{T})$ models as follows.

\subsection{$f(T,\mathcal{T})=T+ \gamma \; g( \mathcal{T})  $ gravity}

With the help of the above quantities, we can obtain the field
equations (modified Friedmann Eq. (\ref{eq10})) as follows
\begin{eqnarray}
3H^2&=& \rho_m -  \frac{\gamma }{2} g(\mathcal{T}) + g_{\mathcal{T}} (\rho_m + p_m) \,\,\,,\label{manuel9}\\
-3H^2-2\dot{H}&=& p_m + \frac{\gamma}{2} g(\mathcal{T})
\,\,,\label{manuel10}
\end{eqnarray}
where $\rho_m$ and $p_m$ denote energy density and pressure
corresponding to DM, respectively. The above equations can also be
written as
\begin{eqnarray}
3H^2&=&\rho_m+\rho_{DE}\label{jonas1}\\
-3H^2-2\dot{H}&=& p_m + p_{DE}\label{jonas2}.
\end{eqnarray}
where
\begin{eqnarray}
\rho_{DE}=-\frac{\gamma }{2} g(\mathcal{T}) + g_{\mathcal{T}}
(\rho_m + p_m), \quad p_{DE}=\frac{\gamma }{2}
g(\mathcal{T}).\label{jonas2''}
\end{eqnarray}
Combining (\ref{jonas1}) and (\ref{jonas2}), one obtains the
following equation
\begin{eqnarray}\label{manuel20}
\rho_{DE} +p_{DE} &= & g_{\mathcal{T}} (\rho_m + p_m) =
g_{\mathcal{T}} \frac{\mathcal{T} \;(1 + \omega_m)}{(1 - 3
\omega_m)}.
\end{eqnarray}
We can also rewrite (\ref{manuel20}) by considering EoS
$p_{DE}=\omega_{DE} \rho_{DE}$ for HDE as
\begin{eqnarray}\label{manuel21}
-2H^2\Omega_{DE} \left(1-\frac{\sqrt{\Omega_{DE}}}{e}\right)=
g_{\mathcal{T}} \frac{\mathcal{T} \;(1 + \omega_m)}{(1 - 3
\omega_m)}\,\,\,.
\end{eqnarray}
For determining $g(T)$ coming from HDE, we assume the Hubble
parameter as follows
\begin{eqnarray}\label{manuel22}
H(t)= h\left(t_s-t\right)^{-\alpha}\,\,\,,
\end{eqnarray}
where $h$ and $\alpha$ appear as positive constants which are taken
as to explain acceleration of the universe. As $t_s$ is future
singularity finite time, such that $t<t_s$. Using (\ref{manuel22})
and (\ref{manuel8}), one has
\begin{eqnarray}\label{manuel23}
\dot{H}=\alpha
h\left[-\frac{T}{6h^2}\right]^{\frac{\alpha+1}{2\alpha}}\,\,\,,
\end{eqnarray}
with which we rewrite Eq.(\ref{manuel21}) as
\begin{eqnarray}\label{manuel24}
\frac{-T}{3} \Omega_{DE}\left(1-\frac{\sqrt{\Omega_{DE}}}{e}\right)=
g_{\mathcal{T}} \frac{\mathcal{T} \;(1 + \omega_m)} {(1 - 3
\omega_m)} \,\,\,.
\end{eqnarray}
The scale factor for (\ref{manuel22}) takes the form
$a(t)=a_0e^{\frac{h(t_s-t)^{1-\alpha}}{\alpha-1}}$. For this scale
factor, the event horizon takes the form
\begin{eqnarray}\label{manuel25}
R_h=a_0e^{\frac{h(t_s-t)^{1-\alpha}}{\alpha-1}}\int_{t}^{t_s}\frac{1}{a_0}
e^{-\frac{h(t_s-\tilde{t})^{1-\alpha}}{\alpha-1}}d\tilde{t}.
\end{eqnarray}
we take  $\alpha=1$ for the sake of simplicity. Then,
Eq.(\ref{manuel25}) becomes
\begin{eqnarray}\label{manuel26}
R_h=\frac{t_s-t}{1+h}\,\,\,,
\end{eqnarray}
from which we get
\begin{eqnarray}\label{manuel27}
\Omega_{DE}=\frac{e^2h^2}{(1+h)^2}\,\,\,,
\end{eqnarray}
and Eq.(\ref{manuel24}) takes the form
\begin{eqnarray}\label{manuel28}
K= g_{\mathcal{T}}\;\mathcal{T}
\end{eqnarray}
where $K$ is a constant depending on $h$ and $e$ as
\begin{eqnarray}\label{manuel29}
K=   \frac{-T}{3} \; \frac{e^2h^2\;(1 - 3 \omega_m)}{(1 +
\omega_m)\;(1+h)^2} \left(1-\frac{h}{1+h}\right)  \,\,\,.
\end{eqnarray}
The Eq. (\ref{manuel28}) gives the following solution
\begin{eqnarray}\label{manuel30}
g(\mathcal{T})=A\; \ln \mathcal{T}^K \,\,\,,
\end{eqnarray}
and the corresponding $f(T,\mathcal{T})$ gravity model according to
HDE is
\begin{eqnarray}\label{manuel31}
f(T,\mathcal{T})=T + A\; \ln \mathcal{T}^K \,\,\,,
\end{eqnarray}
where $K$ and $A$ are constant. Also, the torsion scalar takes the
form $T_0$ \footnote{It is easy to obtain this value through Eq.
(\ref{manuel8}) in term of the initial Hubble parameter $H_0$
according to the observational data} at early time $t_0$. Thus we
have
\begin{eqnarray}\label{manuel32}
\left(\frac{dT}{dt}\right)_{t=t_0}=-12h^2\left(-\frac{T_0}{6h^2}\right)^{\frac{3}{2}}\,\,\,.
\end{eqnarray}
For determining the respective value of $A$, we use the initial
conditions as in $f(R)$ theory of gravity (Wu and Zhu 2008) and
hence the function $g(\mathcal{T})$ must obey the following initial
conditions
\begin{eqnarray}\label{manuel31}
\left(g\right)_{t=t_0}=T_0\,\,\,, \quad\quad
\left(\frac{dg}{dt}\right)_{t=t_0}
=\left(\frac{dT}{dt}\right)_{t=t_0}\,\,\,\,.
\end{eqnarray}
which gives
\begin{eqnarray}\label{manuel32}
\ln A = -\ln \mathcal{T}_0^K \,\,\,.
\end{eqnarray}
Hence, the function $f(T, \mathcal{T})$ has taken the following form
\begin{eqnarray}
f(T, \mathcal{T})=T +\gamma  \ln
\Big(\frac{\mathcal{T}}{\mathcal{T}_0} \Big)^K.
\end{eqnarray}

\subsection{  $f(T,\mathcal{T})=\beta\; \mathcal{T}+ g(T)$ gravity}

For this model, the field equations (\ref{lagran1}) turns out to be
\begin{eqnarray}\label{manuel9'}
3H^2&=& \rho_m--\frac{1}{2}f(T)-6H^2f_T +\frac{2\beta \;  \omega_m
\;\mathcal{T}}{(1-3 \omega_m)},\\\nonumber -3H^2-2\dot{H}&=& p_m +
\frac{1}{2}f(T) + 2\left(3H^2+\dot{H}\right)f_T-24\dot{H}H^2f_{TT} -
\frac{2\beta \;\omega_m \;\mathcal{T}}{(1-3
\omega_m)},\\\label{manuel10'}
\end{eqnarray}
For this model, the field equations can also be written as
\begin{eqnarray}
3H^2&=&\rho_m+\rho_{DE}\,,\label{jonas1'}\\
-3H^2-2\dot{H}&=& p_m + p_{DE}\label{jonas2'}
\end{eqnarray}
with
\begin{eqnarray}
\rho_{DE}&=&-\frac{1}{2}g(T)-6H^2g_T + \frac{2\beta\; \omega_m
\;\mathcal{T}}{(1-3 \omega_m)} \,\,\,,\label{jonas1'}\\
p_{DE}&=&\frac{1}{2}g(T) +
2\left(3H^2+\dot{H}\right)g_T-24\dot{H}H^2g_{TT} - \frac{2\beta
\omega_m \;\mathcal{T}}{(1-3 \omega_m)} \,\,\,.\label{jonas2'}
\end{eqnarray}
By adding (\ref{jonas1'}) and (\ref{jonas2'}), we can obtain
\begin{eqnarray}\label{manuel20'}
\rho_{DE} + p_{DE} = 2\dot{H}g_T-24\dot{H}H^2g_{TT}\,\,\,.
\end{eqnarray}
We can rewrite (\ref{manuel20'}) for this model as follows
\begin{eqnarray}\label{manuel21'}
-2H^2\Omega_{DE}
\left(1-\frac{\sqrt{\Omega_{DE}}}{e}\right)=2\dot{H}g_T-24\dot{H}H^2g_{TT}\,\,\,.
\end{eqnarray}
Using (\ref{manuel8}) and (\ref{manuel22}), we rewrite
Eq.(\ref{manuel21'}) as
\begin{eqnarray}\label{manuel24'}
2Tg_{TT}+g_T-\frac{1}{\alpha h^3}\Omega_{DE}
\left(1-\frac{\sqrt{\Omega_{DE}}}{e}\right)
\left[-\frac{T}{6h^2}\right]^{\frac{\alpha-1}{2\alpha}}=0\,\,\,.
\end{eqnarray}
By making use of (\ref{manuel27}), Eq.(\ref{manuel24'}) takes the
form
\begin{eqnarray}\label{manuel28'}
2Tg_{TT}+g_T+Q=0\,\,\,,
\end{eqnarray}
where $Q$ is a constant depending on $h$ and $e$ as
\begin{eqnarray}\label{manuel29'}
Q=-\frac{e^2h^3}{(1+h)^3}\,\,\,.
\end{eqnarray}
The general solution of (\ref{manuel28'}) is
\begin{eqnarray}\label{manuel30'}
g(T)=-Q T+2\gamma_1 \sqrt{-T}+\gamma_2\,\,\,,
\end{eqnarray}
which also becomes
\begin{eqnarray}\label{manuel31'}
g(T)=\left(1-K\right)T+ 2\gamma_1 \sqrt{-T}+\gamma_2\,\,\,,
\end{eqnarray}
where $\gamma_1$ and $\gamma_2$ are constants which can be found as
follows
\begin{eqnarray}\label{manuel32'}
\left(\frac{dT}{dt}\right)_{t=t_0}=-12h^2\left(-\frac{T_0}{6h^2}\right)^{\frac{3}{2}}\,\,\,.
\end{eqnarray}
With the help of initial conditions, one can get $\gamma_1$ and
$\gamma_2$ as follows
\begin{eqnarray}\label{manuel3'2}
\gamma_1= Q \sqrt{-T_0}\,\,\,,\quad\quad \gamma_2=-KT_0\,\,\,.
\end{eqnarray}
We can then write the explicit expression of $g(T)$ as
\begin{eqnarray}
f(T,\mathcal{T})=\beta\;   \mathcal{T} +  \left(1-Q
\right)T+2Q\sqrt{T_0 T} - QT_0\,\,\,.
\end{eqnarray}

We can observe that when HDE contribution is almost null, i.e.
($e=0$,\,or $Q=0$ and $\beta=0$), $f(T)=T$ (the teleparallel gravity
equivalent to GR). For $\beta=0$, the above result corresponds to
$f(R)$ gravity (Nojiri and Odintsov 2006b), This allows us to tell
our model provides a general aspect.

\section{Conclusion}

The reconstruction of $f(T, \mathcal T)$ gravity for HDE model has
been presented in this paper. Two sepcific models has been adopted
for our work such as $f(T, \mathcal T) = T + \gamma g(\mathcal T)$
(a correction to the teleparallel action depending on the matter
content) and $f(T,\mathcal T) =\beta \mathcal T + g(T)$ (a simple
$\mathcal T$-linear correction to the class of $f(T)$ theories). In
these cases, we can recover teleparallel gravity by setting
$g(\mathcal T)=0$ or $g(T)=0$. We have obtained the equation of
motion for the flat FRW universe. We found that the differential
equations were solved analytically by use of the initial conditions.
Thus, we have solved these differential equations for $g(T)$ and
obtain $g$ and corresponding $f$. The introduced constants
$\gamma_1$ and $\gamma_2$ has been determined on the basis
of initial conditions as mentioned in $f(R)$ gravity.\\

{\bf Acknowledgement}:  Ines G. Salako thanks IMSP  for hospitality
during the elaboration of this work. Surajit Chattopadhyay sincerely
acknowledges financial support under DST Project SR/FTP/PS-167/2011
and the visiting associateship of IUCAA,
Pune, India.\\

\end{document}